# A Note on a Generalization of Sherman-Morrison-Woodbury formula

Milan Batista

University of Ljubljana, Faculty of Maritime Studies and Transport

Pot pomorščakov 4, 6320 Portorož, Slovenia, EU

milan.batista@fpp.edu

## Abstract

The article presents a generalization of Sherman-Morrison-Woodbury (SMW) formula for the inversion of a matrix of the form $A + \sum_{k=1}^{N} U_k V_k^T$.

In the author recent paper ([1]) the SMW formula ([2]) was used to show how the cyclic block tridiagonal and cyclic block pentadiagonal systems can be solved. The solution of cyclic block pentadiagonal system depends on use of Eq 30 in [1] which generalize SMW formula to the case $k = 2$. However this can be, as will be shown below, generalized to any $k = N$. In what follows all the matrices are assumed to be nonsingular.

**Theorem**. If $A$ is $n \times n$ matrix and $U_k$ and $V_k$ $(k = 1,..,N)$ are $n \times m$ matrices then

$$\left( A + \sum_{k=1}^{N} U_k V_k^T \right)^{-1} = A^{-1} - A^{-1} [U_1, U_2, ..., U_N] M^{-1} [V_1, V_2, ..., V_N]^T A^{-1} \quad (1)$$

where $M$ is $Nm \times Nm$ matrix given by

$$M = \begin{bmatrix} I_{m \times m} + V_1^T A^{-1} U_1 & V_1^T A^{-1} U_2 & \cdots & V_1^T A^{-1} U_N \\ V_2^T A^{-1} U_1 & I_{m \times m} + V_2^T A^{-1} U_2 & \cdots & V_2^T A^{-1} U_N \\ \vdots & & & \vdots \\ V_N^T A^{-1} U_1 & V_N^T A^{-1} U_2 & \cdots & I_{m \times m} + V_N^T A^{-1} U_N \end{bmatrix} \quad (2)$$

*Proof*: Introducing $n \times Nm$ matrices $B = [U_1, U_2, ..., U_N]^T$ and $C = [V_1^T, V_2^T, ..., V_N^T]^T$ one finds that $BC^T = \sum_{k=1}^{N} U_k V_k^T$. By SMW formula on has

$$(A + BC^T)^{-1} = A^{-1} - A^{-1}B(I_{Nm \times Nm} + C^T A^{-1} B)^{-1} C^T A^{-1} \qquad (3)$$

Now it has to be shown that $M = I_{Nm \times Nm} + C^T A^{-1} B$. By using definition of $B$ and $C$ one gets by direct computation

$$I_{Nm \times Nm} + C^T A^{-1} B = I_{Nm \times Nm} + [V_1^T, V_2^T, ..., V_N^T]^T [A^{-1}U_1, A^{-1}U_2, ..., A^{-1}U_N] \qquad (4)$$

which in next step of calculation yield matrix (2).

□□□

**Corollory.** Let $A = I$ the $n \times n$ identity matrix. Then

$$\left(I + \sum_{k=1}^{N} U_k V_k^T \right)^{-1} = I - [U_1, U_2, ... U_N] M^{-1} [V_1, V_2, ..., V_N]^T \qquad (5)$$

where

$$M = \begin{bmatrix} I_{m \times m} + V_1^T U_1 & V_1^T U_2 & \cdots & V_1^T U_N \\ V_2^T U_1 & I_{m \times m} + V_2^T U_2 & \cdots & V_2^T U_N \\ \vdots & \vdots & & \vdots \\ V_N^T U_1 & V_N^T U_2 & \cdots & I_{m \times m} + V_N^T U_N \end{bmatrix} \qquad (6)$$

**Acknowledge.** I wish to thanks Yi Yu from wpi.edu for notice flaw in Eq 1 and 6 in the previous version of the paper.